\documentclass[aps,pre,twocolumn,groupedaddress,nopacs]{revtex4}
\usepackage{graphicx}
\usepackage{dcolumn}
\usepackage{bm}
\usepackage{amssymb}
\usepackage{color}

\begin{document}


\title{Shape transitions of high-genus fluid vesicles} 

\author{Hiroshi Noguchi}
\email[]{noguchi@issp.u-tokyo.ac.jp}
\affiliation{
Institute for Solid State Physics, University of Tokyo,
 Kashiwa, Chiba 277-8581, Japan}


\begin{abstract}
The morphologies of genus-2 to -8 fluid vesicles are studied
by using dynamically triangulated membrane simulations with area-difference elasticity.
It is revealed that the alignments of the membrane pores alter the vesicle shapes and 
the types of shape transitions for  the genus $g \ge 3$.
At a high reduced volume, a stomatocyte with a circular alignment of $g+1$ pores
continuously transforms into a discocyte with a line of $g$ pores with increasing intrinsic area difference.
In contrast, at a low volume, a stomatocyte transforms into a ($g+1$)-hedral shape 
and subsequently exhibits a discrete phase transition to a discocyte.
\end{abstract}


\maketitle

\section{Introduction}

Cell organelles have a variety of morphologies.
In some  organelles, lipidic necks or pores connect biomembranes
such that they hold a nonzero genus.
For example, a nuclear envelope consists of two bilayer membranes
 connected by many lipidic pores, which is supported by protein complexes.
The outer  nuclear membrane is also connected  by the endoplasmic reticulum,
which consists of tubular networks and flat membranes.
The fusion and fission of the membrane tubes change their genus.
It is important to understand the topology dependence of their morphologies.

Since lipid bilayer membranes are basic components of biomembranes,
lipid vesicles in a fluid phase are considered a simple model of cells and organelles.
The morphologies of genus-0 vesicles have been intensively studied 
experimentally and theoretically~\cite{canh70,helf73,svet89,lipo95,seif97,hota99,saka12,kahr12a,saka14}.
A red-blood-cell shape (discocyte) as well as prolate and stomatocyte,
can be reproduced by minimization of the bending energy with area and volume constraints.
Other shapes such as a pear and branched tubes are obtained by the addition of 
 spontaneous curvature or area-difference elasticity (ADE)~\cite{svet89,lipo95,seif97}.
In particular, the ADE model can reproduce experimentally observed liposome shapes  very well~\cite{saka12}.
In contrast to the genus-0 vesicles,
vesicles with a nonzero genus have been much less explored.

Vesicle shapes with the genuses $g=1$ and $g=2$ were studied in the 1990s \cite{zhon90,seif91a,four92,juli93,juli93a,juli96,mich94,mich95}.
For $g=1$, a conformational degeneracy was found in the ground state of the bending energy,
where the vesicles can transform their shapes if the vesicle volume is allowed to freely vary~\cite{seif91a}.
For $g\ge 2$, a conformational degeneracy is obtained even with a fixed volume~\cite{juli93a,juli96,mich95}.
Phase diagrams of genus-1 and genus-2 vesicles were  constructed by J{\"u}licher et al.~\cite{juli93,juli93a,juli96}
for symmetric shapes.
Recently, we found that nonaxisymmetric shapes such as elliptic and handled discocytes also exist in equilibrium for genus-1 vesicles~\cite{nogu15a}.

Vesicles with $g\gg 1$ were observed for polymersomes, and
the budding of hexagonally arranged pores was investigated~in Ref.~\cite{halu02}.
To our knowledge, this paper is the only previous study of the shape transition of vesicles with $g\ge 3$.
Thus, vesicle shapes for $g\ge 3$ have been little explored.
In this study, we investigate the vesicle shapes for $2 \leq g \leq 8$ by using dynamically triangulated membrane simulation with the ADE model.
In particular, we focus on the genus-5 vesicles,
where various shapes including cubic and five-fold symmetrical shapes are formed.
We will show that the arrangements of the pores remarkably change the shape transitions.

\section{Simulation Model and Method}\label{sec:method}

The morphologies of fluid vesicles are simulated by 
 a dynamically triangulated surface method \cite{gomp04c,nogu09,nogu15a}.
Since the details of the potentials are described in Ref.~\cite{nogu15a} and
the general features of the triangulated membrane can be found in Ref.~\cite{gomp04c},
 the membrane model is briefly described here.
A vesicle consists of
$4000$ vertices with a hard-core excluded volume of diameter $\sigma_0$. 
The maximum bond length is $\sigma_1=1.67\sigma_0$.
The volume $V$ and surface area $A$ are maintained by harmonic potentials $U_{\rm {V}}= (1/2)k_{\rm {V}}(V-V_{\rm 0})^2$ and
$U_{\rm {A}}= (1/2)k_{\rm {A}}(A-A_{\rm 0})^2$ with  $k_{\rm {V}}=4k_{\rm B}T$ and $k_{\rm {A}}=8k_{\rm B}T$,
where $k_{\rm B}T$ is the thermal energy.
The deviations in the reduced volume $V^*=V/(4\pi/3){R_{\rm A}}^{3}$ from the target values are less than $0.1$\%,
where $R_{\rm A}=\sqrt{A/4\pi}$.
A Metropolis Monte Carlo (MC) method is used for vertex motion
and reconnection of the bonds (bond flip).

The bending energy of a single-component fluid vesicle is given by~\cite{canh70,helf73}
\begin{equation}
U_{\rm {cv}} =  \int  \frac{\kappa}{2}(C_1+C_2)^2   dA,
\label{eq:cv}
\end{equation}
where $C_1$ and $C_2$ are the principal curvatures at each point 
in the membrane. The coefficient $\kappa$ is the bending rigidity.
The spontaneous curvature and Gaussian bending energy
are not taken into account since the spontaneous curvature
 vanishes for a homogeneous bilayer membrane
and the integral over the Gaussian curvature $C_1C_2$ is invariant for a fixed topology.

In the ADE model, the ADE energy $U_{\rm {ADE}}$ is added as follows \cite{seif97,lipo95,svet89}:
\begin{equation}
U_{\rm {ADE}} =  \frac{\pi k_{\rm {ade}}}{2Ah^2}(\Delta A - \Delta A_0)^2.
\label{eq:ade}
\end{equation}
The areas of the outer and inner monolayers of a bilayer vesicle
differ by $\Delta A= h \int (C_1+C_2) dA$,
where $h$ is the distance between the two monolayers.
The area differences are normalized by a spherical vesicle as
$\Delta a =\Delta A/8\pi h R_{\rm A}$ and $\Delta a_0 = \Delta A_0/8\pi h R_{\rm A}$
 to display our results.
The spherical vesicle with $\Delta a_0=0$ has $\Delta a =1$ and $U_{\rm {ADE}} =8\pi^2 k_{\rm {ade}}$.
The mean curvature at each vertex is discretized using dual lattices \cite{nogu15a,gomp04c,itzy86,nogu05}.

In the present simulations,
we use  $\kappa=20k_{\rm B}T$ and $k_{\rm {ade}}^* = k_{\rm {ade}}/\kappa=1$.
These are typical values for phospholipids~\cite{seif97,saka12}.
Most of the simulations are performed with the bending and ADE potentials under the volume and area constraints.
In some of the simulations,  $k_{\rm {ade}}^*=0$ and $k_{\rm {V}}=0$ are employed 
in order to simulate the vesicles without the ADE energy and volume constraints,
respectively.
In the long time limit, the area difference is relaxed to $\Delta a=\Delta a_0$, 
although it does not  occur on a typical experimental time scale.
The canonical MC simulations of the ADE model are performed from different initial conformations for various values of $V^*$ and $\Delta a_0$.
To obtain the thermal equilibrium states,
a replica exchange MC (REMC) method \cite{huku96,okam04} with $8$ to $24$ replicas is employed for the genus-5 vesicles.
Different replicas have different values of $\Delta a_0$ or $V_0$ and neighboring replicas exchange them by the Metropolis method.

\begin{figure}
\includegraphics{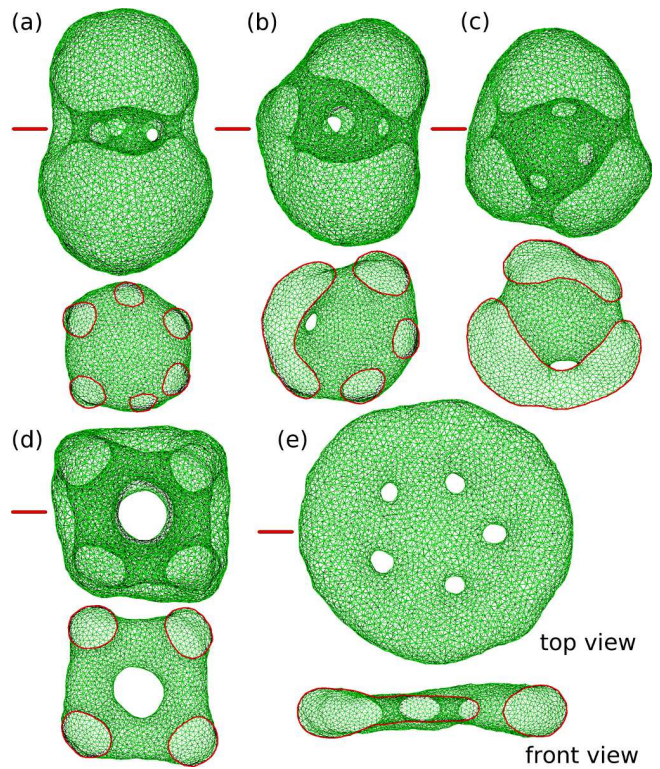}
\caption{
Snapshots of genus-5 vesicles
obtained by the simulation (a) without the volume constraint at $k_{\rm {ade}}^*=0$,
 (b),(c) with the volume constraint at $k_{\rm {ade}}^*=0$,
and (d), (e) with the volume constraint at $k_{\rm {ade}}^*=1$ and $\Delta a_0=1.45$.
(a) Circular-cage stomatocyte at $V^*=0.63$ and $\Delta a=0.98$.
(b) Stomatocyte at $V^*=0.54$ and $\Delta a=0.8$.
(c) Spherical stomatocyte at $V^*=0.5$ and $\Delta a=0.74$.
(d) Cube at $V^*=0.4$ and $\Delta a=1.41$.
(e) Discocyte at $V^*=0.4$ and $\Delta a=1.32$.
The top and front views are shown.
In the front view, the vesicles are cut along a plane shown as a (red) straight line on the left side in the top view.
Their front halves are removed and
the cross sections are indicated by the thick (red) lines.
}
\label{fig:snap_stov4}
\end{figure}

\begin{figure}
\includegraphics{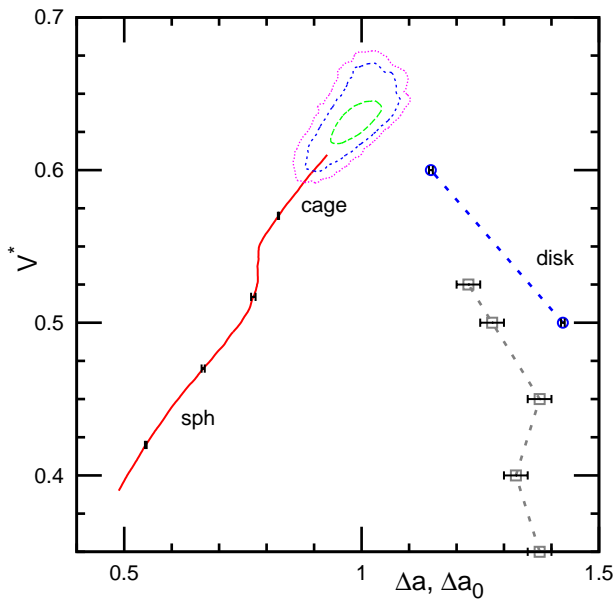}
\caption{
Phase diagram of genus-5 vesicles.
The (red) solid line represents the mean area difference $\langle \Delta a \rangle$
obtained from the simulation with the volume constraint at $k_{\rm {ade}}^*=0$.
The contour lines of the probability distribution $P(V^*,\Delta a)=0.00001,0.0001,0.001$ 
are obtained by the simulation without the volume constraint at $k_{\rm {ade}}^*=0$.
The circles and squares with the dashed lines represent the phase boundaries of $\Delta a_0$ for
stable and metastable discocytes, respectively.
The error bars are shown at several data points for $\langle \Delta a \rangle$
and at all data points for the phase boundaries.
}
\label{fig:phase}
\end{figure}

\section{Genus-5 vesicles}\label{sec:genus5}

\begin{figure}
\includegraphics{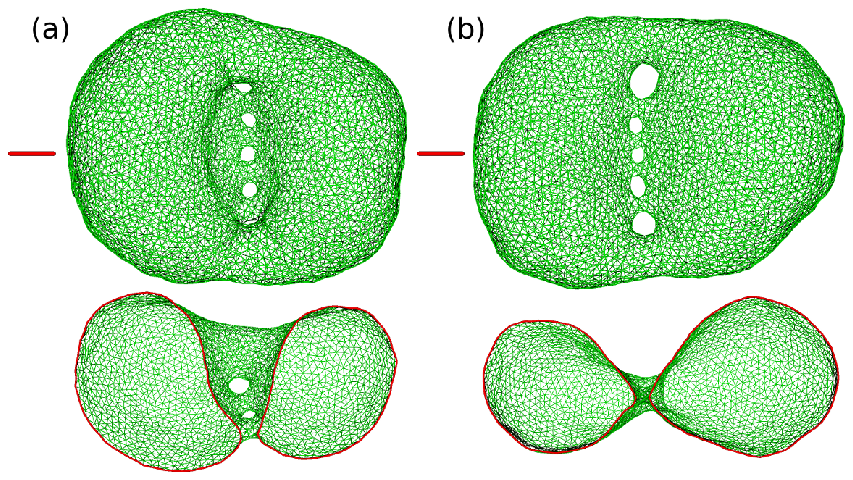}
\includegraphics{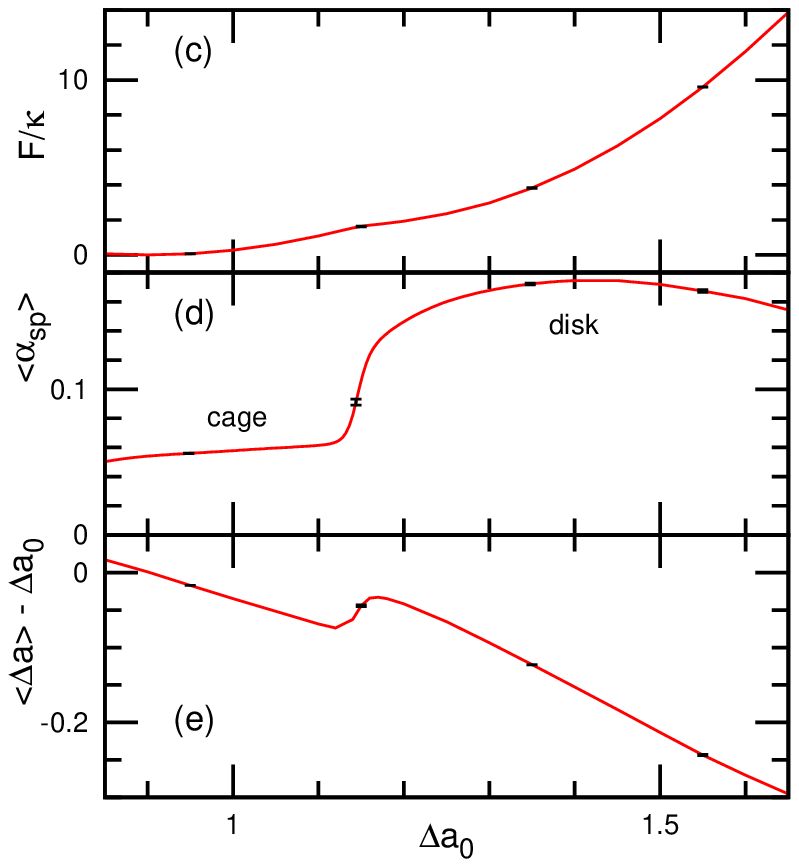}
\caption{
Dependence of the vesicle shapes on $\Delta a_0$ for  genus-5 vesicles at $V^*=0.6$ and $k_{\rm {ade}}^*=1$.
(a),(b) Snapshots at $\Delta a_0=$ (a) $1.16$ and (b) $1.45$.
(c) Free-energy profile $F$.
(d) Mean asphericity $\langle \alpha_{\rm {sp}} \rangle$.
(e) Mean area difference $\langle \Delta a \rangle$ compared to the intrinsic area difference 
$\Delta a_0$.
The error bars are shown at several data points.
}
\label{fig:v6}
\end{figure}

\begin{figure}
\includegraphics{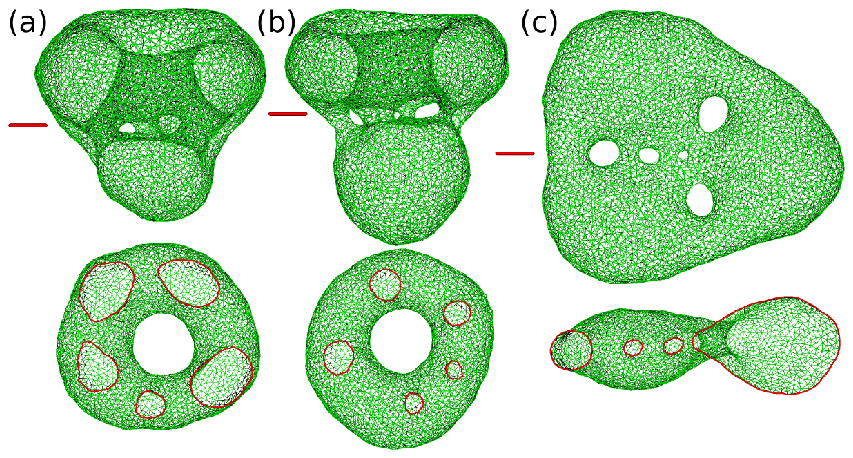}
\includegraphics[width=7.5cm]{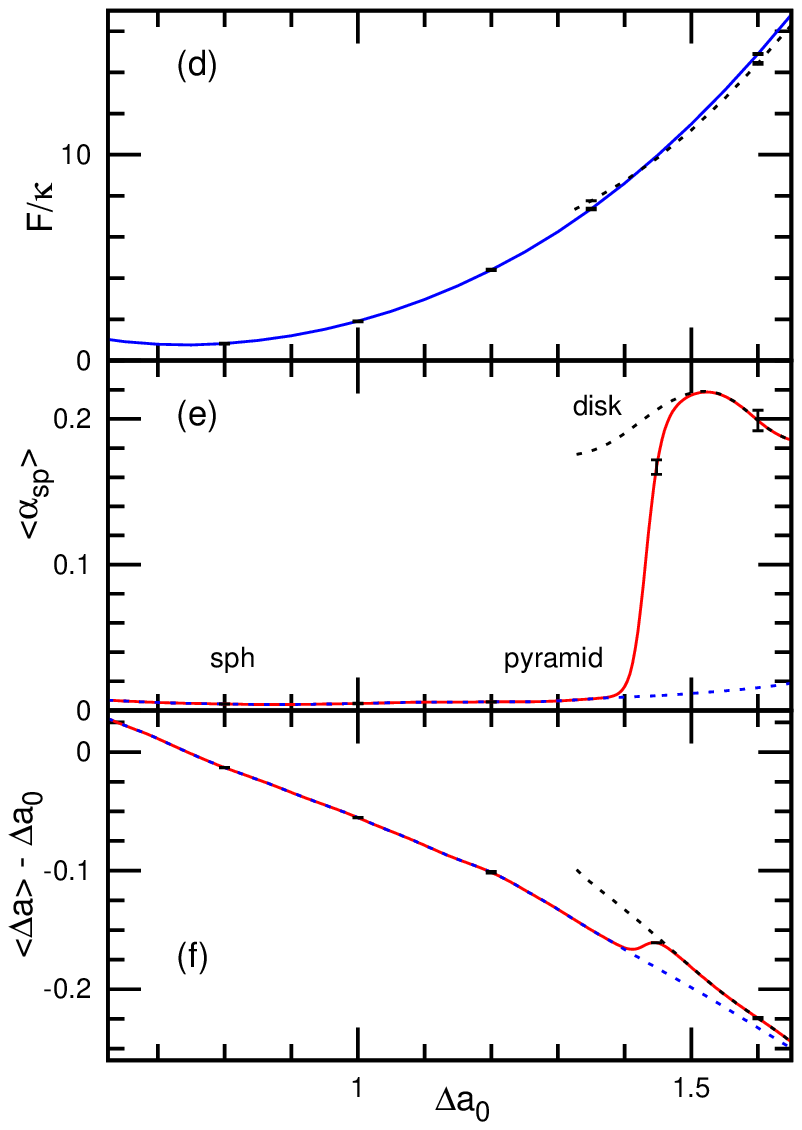}
\caption{
Dependence of the vesicle shapes on $\Delta a_0$ for  genus-5 vesicles at $V^*=0.5$ and $k_{\rm {ade}}^*=1$.
(a) Snapshot of a pentagonal pyramid at $\Delta a_0=1.3$.
(b) Snapshot of a concave pentagonal pyramid at $\Delta a_0=1.6$.
(c) Snapshot of a discocyte at $\Delta a_0=1.6$.
(d) Free-energy profile $F$.
The (blue) solid line represents $F$ of the spherical and pyramidal stomatocytes;
the (black) dashed line represents $F$ of the discocyte.
(e) Mean asphericity $\langle \alpha_{\rm {sp}} \rangle$.
(f) Mean area difference $\langle \Delta a \rangle$ compared to the intrinsic area difference 
$\Delta a_0$.
The solid lines in (e) and (f) represent the data in equilibrium
and the dashed lines represent the data averaged for either pyramid or discocyte.
The error bars are shown at several data points.
}
\label{fig:v5}
\end{figure}

We intensively investigate vesicles with the genus $g=5$ (Figs.~\ref{fig:snap_stov4}--\ref{fig:v5}).
We categorize the vesicle shapes as stomatocyte and discocyte.
A spherical invagination with a narrow neck into the inside of a spherical vesicle
is a typical stomatocyte.
Here, we also include a nonspherical invagination [see Fig.~\ref{fig:snap_stov4}(a)] 
and an invagination with wide necks [see Fig.~\ref{fig:snap_stov4}(d)] into the stomatocyte.
On the discocyte, five pores can be aligned in circular, straight, or branched lines 
[see Figs.~\ref{fig:snap_stov4}(e), \ref{fig:v6}(b), and \ref{fig:v5}(c)].

When the volume constraint and the ADE potential are removed ($k_{\rm {V}}=0$ and $k_{\rm {ade}}^*=0$),
the vesicle shape is determined by the bending energy with the topological constraint.
As mentioned in the Introduction,
the lowest bending-energy states of the vesicles with $g\geq 1$ are not a single conformation
but the vesicles transform shapes without changing the bending energy~\cite{juli93a,juli96,mich95}.
At $g=5$, the lowest energy states are a circular-cage stomatocyte,
where $g+1$ pores are aligned in a circular line 
and the pore size largely fluctuates [see Fig.~\ref{fig:snap_stov4}(a)].
The vesicle shapes are distributed around $V^*\simeq 0.63$ and $\Delta a\simeq 1$
as shown in Fig.~\ref{fig:phase}.
Note that the simulations are performed at a finite temperature ($\kappa=20k_{\rm B}T$)
so that the obtained shapes contain thermally excited states around the lowest energy states.

Stomatocytes are stable shapes with the volume constraint for $k_{\rm {ade}}^*=0$.
A valley in the free-energy landscape in the $V^*$--$\Delta a$ space is extended from the lowest energy state 
to low reduced volumes along the solid (red) line in Fig.~\ref{fig:phase}.
This line is calculated using the REMC method for $V^*$ with $k_{\rm V}=0.0001$ without the ADE potential.
With decreasing $V^*$, it is found that the vesicle transforms from the circular-cage shape
to a spherical stomatocyte [see Fig.~\ref{fig:snap_stov4}(c)],
where $g+1$ pores are distributed on the vesicle surface.
This transformation occurs as a continuous change.
In the transient region at  $0.52 \lesssim V^*\lesssim 0.55$,
the vesicle has intermediate shapes, in which
the positions of one or two pores often deviate from a plane  [see Fig.~\ref{fig:snap_stov4}(b)].
This shape transformation in the stomatocytes is characteristic for vesicles with $g\geq 3$
but is not obtained for $g\leq 2$.
Thus, the arrangement of the pores appears as a new factor for the high-genus vesicles.

Next, we describe the stomatocyte--discocyte transition
as a function of $\Delta a_0$  using the ADE model.
It drastically changes above or below the critical reduced volume $V^*\simeq 0.54$
of the change in the pore arrangement on the stomatocyte.
Figure~\ref{fig:v6} shows the free energy and shapes of the vesicles at $V^*=0.6$ calculated 
by the REMC simulation.
The lowest free-energy state is  the circular-cage stomatocyte at $\Delta a_0=0.9$ [see Fig.~\ref{fig:v6}(c)]. 
We takes this to be the origin of the free energy, $F=0$, in this study.
With increasing $\Delta a_0$, one of the pores in the cage stomatocyte
gradually opens [see Fig.~\ref{fig:v6}(a)],
and subsequently, a discocyte with a straight line of $g$ pores is formed [see Fig.~\ref{fig:v6}(b)].

The vesicle shapes are quantified by a shape parameter, asphericity $\alpha_{\rm {sp}}$,
and the area difference $\Delta a$. 
The asphericity $\alpha_{\rm {sp}}$ is defined as \cite{rudn86}
\begin{equation}
\alpha_{\rm {sp}} = \frac{({\lambda_1}-{\lambda_2})^2 + 
  ({\lambda_2}-{\lambda_3})^2+({\lambda_3}-{\lambda_1})^2}{2 (\lambda_1+\lambda_2+\lambda_3)^2},
\end{equation}
where ${\lambda_1} \leq {\lambda_2} \leq {\lambda_3}$ are the 
eigenvalues of the gyration tensor of the vesicle.
The asphericity is the degree of deviation from a spherical 
shape:  $\alpha_{\rm {sp}} = 0$ for spheres, $\alpha_{\rm {sp}}=1$ 
for thin rods, and $\alpha_{\rm {sp}}=0.25$ for thin disks \cite{nogu05,nogu15a}.
The spherical stomatocyte and discocyte have $\alpha_{\rm {sp}} \simeq 0$ and $0.2$, respectively.
The circular-cage stomatocyte has $\alpha_{\rm {sp}} \simeq 0.05$.
As the stomatocyte transforms into the discocyte,
$\alpha_{\rm {sp}}$ and $\Delta a$ changes abruptly.
The transition point ($\Delta a_0=1.146\pm0.004$) is estimated 
from the maximum slope of the $\alpha_{\rm sp}$ curves in Fig.~\ref{fig:v6}(d).
When the discocyte vesicle transforms into the open stomatocyte,
the mirror symmetry breaks,
and the slope of $F(\Delta a_0)$ changes.
This is a second-order type of transition, but it is rounded by the thermal fluctuations.
This characteristic of the transition is the same as the stomatocyte--discocyte transition at $g=0$ and $1$.

The stomatocyte--discocyte transformation becomes a discrete transition
  below the critical reduced volume $V^*\simeq 0.54$.
We obtained the coexistence of the stomatocyte and discocyte using canonical MC simulations
in the right region of the dashed (gray) line with squares
in Fig.~\ref{fig:phase}. The final shapes are determined by hysteresis from the initial conformations.
For example, cubic and discocyte vesicles are obtained at $V^*=0.4$ and $\Delta a_0=1.45$ 
[see Figs.~\ref{fig:snap_stov4}(d) and (e)].
However, the canonical simulation cannot determine which shape is more stable.
Both shapes can be maintained without transformation into the other shape in this region.
When we use the REMC simulation for $\Delta a_0$,
the exchange between the stomatocyte and the discocyte does not occur in equilibrium owing to a large free-energy barrier.
To overcome such an energy barrier, one more order parameter is often added 
and the REMC or alternative generalized ensemble method is performed 
in a partially or fully two-dimensional parameter space. 
Previously, we employed the asphericity as an additional order parameter for a constant radius of gyration
in order to overcome the energy barrier between the two free-energy valleys
of the discocyte and prolate of a genus-0 vesicle~\cite{nogu05}.
For the current transition, however, we did not succeed with similar strategies.
Therefore, we take a detour via the stomatocyte--discocyte continuous transition at $V^*=0.6$.
This type of detour has been used in free-energy calculations \cite{smir15,muel08}.
For liquid--gas phase transitions the barrier can be avoided  via supercritical fluids.
For membranes, an external order field was used to investigate the formation energy of a fusion intermediate \cite{smir15}.

Here, we use three REMC simulations to estimate the free-energy difference $\Delta F_{\rm {v5}}$  between the stomatocyte at $\Delta a_0=0.75$ 
and the discocyte at $\Delta a_0=1.55$ for $V^*=0.5$.
First, the free-energy difference between the stomatocytes at $V^*=0.5$ and $0.6$
is calculated as $\Delta F_{\rm {st}}/k_{\rm B}T=15.1 \pm 0.1$ from the REMC simulation used 
for calculating the solid line in Fig.~\ref{fig:phase}.
Second, the difference  $\Delta F_{\rm {v6}}/k_{\rm B}T=192.1 \pm 0.1$ between the stomatocyte at $\Delta a_0=0.9$ (the free-energy minimum for $k_{\rm {ade}}=0$) 
and the discocyte at $\Delta a_0=1.55$  for $V^*=0.6$
is calculated as shown in Fig.~\ref{fig:v6}(c).
Third, the difference  $\Delta F_{\rm {dis}}/k_{\rm B}T=62.5 \pm 0.1$ between the discocytes at $V^*=0.5$ and $0.6$ for $\Delta a_0=1.55$
is calculated from the REMC simulation for $V^*$ with  $k_V=0.0001$ and $\Delta a_0=1.55$.
Hence, the free-energy difference at $V^*=0.5$ is calculated 
as $\Delta F_{\rm {v5}}= \Delta F_{\rm {v6}}+\Delta F_{\rm {dis}} - \Delta F_{\rm {st}} = 239.5k_{\rm B}T \pm 0.3k_{\rm B}T$.
We simulate the stomatocyte and discocyte at $V^*=0.5$ separately using the REMC simulations
[see the lower (blue) and upper (black) dashed lines in Figs.~\ref{fig:v5}(e) and (f)]
and obtain the equilibrium states by averaging them with the weight $\exp(-\Delta F/k_{\rm B}T)$ as shown in Fig.~\ref{fig:v5}(d).
This calculation clarifies that the stomatocyte--discocyte transformation is a discrete transition at  $\Delta a_0=1.146 \pm 0.004$ for $V^*=0.5$.
[see the solid (red) lines in Figs.~\ref{fig:v5}(e) and (f)].

Next, we investigate the vesicle shapes in detail for $V^*=0.5$.
With an increase in $\Delta a_0$, the spherical stomatocyte transforms into a  pentagonal pyramid [see Fig.~\ref{fig:v5}(a)].
With a further increase, 
the side faces of the pyramid become concave,
and the vesicle shape is represented by
a circular toroid connected to a sphere via five narrow necks  [see Fig.~\ref{fig:v5}(b)].
This concave necked shape prevents the opening of the bottom pore into a discocyte.
On the discocytes, pores are aligned in circular, straight, or branched lines 
[see Figs.~\ref{fig:snap_stov4}(e), \ref{fig:v6}(b), and \ref{fig:v5}(c)].
At $V^*=0.5$, these three alignments coexist.
As  $V^*$ decreases and increases, the circular and straight pore alignments appear more frequently,
and the circular and straight alignments only exist for $V^*=0.4$ and $0.6$, respectively.

\begin{figure}
\includegraphics{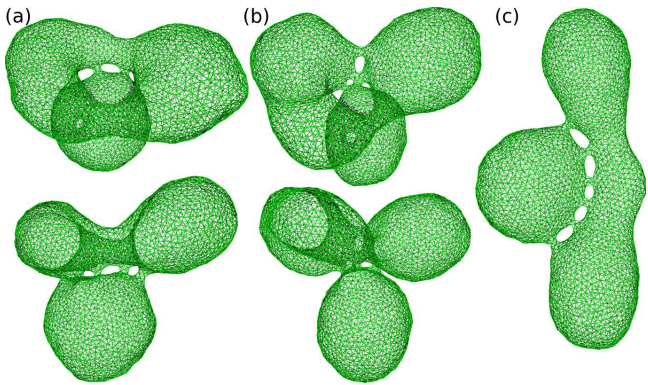}
\caption{
Snapshots of the genus-5 vesicles  at $V^*=0.5$ and $k_{\rm {ade}}^*=1$.
(a),(b) Budded stomatocytes at $\Delta a_0=$ (a) $2$ and (b) $2.2$.
The bird's eye and front views are shown.
(c) Budded discocyte at $\Delta a_0=2.2$.
}
\label{fig:bud}
\end{figure}

\begin{figure}
\includegraphics{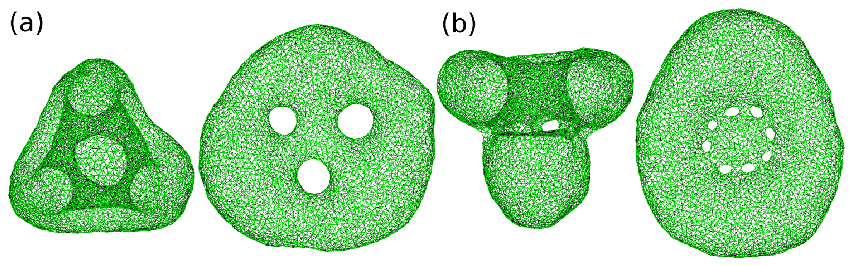}
\includegraphics{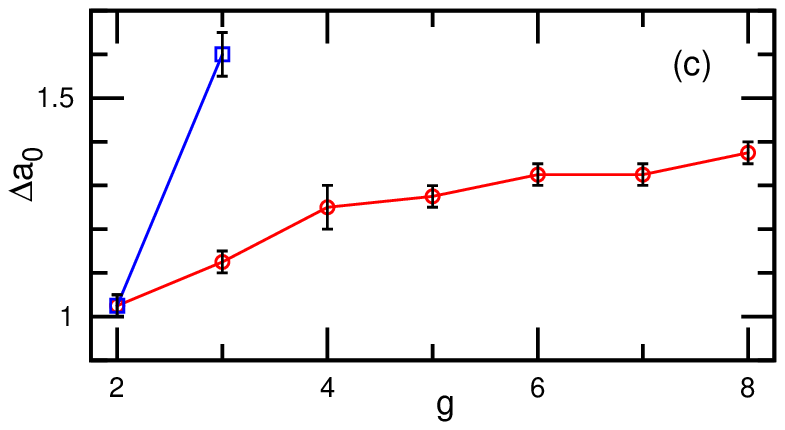}
\caption{
Genus $g$ dependence of the vesicle shapes  at $V^*=0.5$ and $k_{\rm {ade}}^*=1$.
(a) Snapshots of genus-3 vesicles at $\Delta a_0=1.5$.
(b) Snapshots of genus-8 vesicles at $\Delta a_0=1.5$.
(c) Upper ($\Box$) and lower ($\circ$) boundaries of the coexistence regions
as a function of the genus $g$.
The error bars are shown at all data points.
}
\label{fig:disn}
\end{figure}

Our simulations have revealed that the stomatocyte--discocyte transformation changes
from a continuous to a discrete transition as the stomatocyte changes from the circular-cage to the spherical shape.
The crucial difference between these two transitions is the change in the pore alignments.
During the circular-cage to discocyte transformation, the pore alignment does not change 
[see Figs.~\ref{fig:snap_stov4}(a) and \ref{fig:v6}(a),(b)].
In contrast, the transformation from the cubic to discoidal shapes at $V^*=0.4$ is accompanied by the realignment
of the pores from cubic symmetry to five-fold symmetry [see Figs.~\ref{fig:snap_stov4}(d) and (e)].
The transformation from the pyramidal to the discoidal shape at $V^*=0.5$ is not accompanied by pore realignment,
but the circular alignment of five pores stabilizes the half-open stomatocyte  [see Fig.~\ref{fig:v5}(b)].
Thus, the pore alignments are the origin of the discrete transitions.

At $\Delta a_0 \gtrsim 2$, outward buddings occur for both stomatocytes and discocytes.
Since the buds are connected by multiple membrane necks,
the bud shapes are not always spherical.
For the stomatocytes,
the circular-toroidal and spherical compartments are divided and the toroidal compartment elongates at $\Delta a_0 = 2$
[see Fig.~\ref{fig:bud}(a)].
With increasing $\Delta a_0$, budding occurs in narrow regions of the toroid
[see Fig.~\ref{fig:bud}(b)].
Discocytes become divided along the pore alignments 
[see Fig.~\ref{fig:bud}(c)].
With further increasing $\Delta a_0$, more buds are formed.
These budded shapes coexist in a wide range of $\Delta a_0$ because of the free-energy barrier for 
the formation and removal of the narrow necks.

\section{Dependence on Genus}\label{sec:genuss}

The discrete transitions from stomatocytes to discocytes are
obtained for $g\geq 3$ (see Fig.~\ref{fig:disn}).
The features described in the previous section for $g=5$ are general for $g\geq 3$.
A $g$-fold pyramid-shaped vesicle coexists with a discocyte
[see the triangular and concave octagonal pyramids in Figs.~\ref{fig:disn}(a) and (b)].
For $g=3$, stomatocytes transform into discocytes at $\Delta a_0 \gtrsim 1.6$ for $V^*=0.5$.
For  $g\geq 4$, the coexistence region has no upper  $\Delta a_0$ limit. 
At $\Delta a_0 \gtrsim 2$, the vesicles exhibit budding, as for $g=5$.

\section{Summary}
\label{sec:sum}

We revealed that the stomatocyte--discocyte transformation 
is a discrete shape transition at low reduced volumes $V^*$ for $3 \leq g\leq 8$,
while it is a continuous transformation at high $V^*$.
This discreteness is caused by the alignment of $g+1$ pores in the stomatocytes.
At high $V^*$, the pores are aligned in a circular line, while
the pores are distributed on the entire surface at low $V^*$.
As the intrinsic area difference $\Delta a_0$ increases,
this spherical stomatocyte transforms into polyhedral shapes.
The transformations of these vesicles from polyhedral to discoidal shapes are the first-order transition.

We found a continuous transformation from circular-cage to spherical stomatocytes.
However, for vesicles with $g\gg 1$, this transformation may be a discrete transition
since the symmetry of the pores is changed.
The phase behavior at $g\gg 1$ is an interesting problem for further studies.

Polyhedral vesicles are formed under various conditions:
gel-phase membranes~\cite{sack94},  phase-separated membranes~\cite{veat03,gudh07,hu11,nogu12a},
the assembly of protein rods~\cite{nogu15b},
and fluid membranes with the accumulation of specific lipids or defects on the polyhedral edges or vertices~\cite{dubo01,nogu03,hase10}.
However, such vesicles typically have multiple (meta-)stable shapes 
and it is difficult to produce only a specific type of polyhedral shape.
For high-genus vesicles, a polyhedral vesicle has $g+1$ faces.
The vesicles can only form a triangular pyramid for $g=3$ and a cube and pentagonal pyramid  for $g=5$.
Thus, the possible polyhedra are limited and the shape can be controlled by $V^*$ and $\Delta a_0$.
Recently, the assembly and packing of polyhedral objects have received growing attentions \cite{dama12}.
Cubic and other polyhedral vesicles can be interesting building blocks,
since their shapes are deformable and can be controlled by the osmotic pressure.

A mitochondrion consists of two bilayer membranes.
The inner membrane has a much larger surface area than the outer one
and forms many invaginations called cristae.
As a model of such a confinement of an outer vesicle,
the morphology of a genus-0 vesicle under spherical confinement has been recently studied \cite{kahr12a,saka14}.
The confinement induces various shapes such as double and quadruple stomatocytes, 
a slit vesicle, and vesicles of two or three compartments.
Very recently, Bouzar et al. reported that the confinement transforms axisymmetric toroids into asymmetric shapes 
for genus-1 vesicles~\cite{bouz15}.
For higher-genus vesicles, the confinement should similarly stabilize the stomatocytes with respect to the discocytes
since the stomatocytes are more compact.

\begin{acknowledgments}
This work is supported by KAKENHI (25400425) from
the Ministry of Education, Culture, Sports, Science, and Technology of Japan.
\end{acknowledgments}

\end{document}